\documentclass[10pt,twocolumn]{article}
\usepackage[utf8]{inputenc}
\usepackage[T1]{fontenc}
\usepackage{mathptmx}
\usepackage[a4paper,top=18mm,bottom=18mm,left=20.5mm,right=16.5mm,columnsep=6.5mm]{geometry}
\usepackage{amsmath,amssymb,amsthm}
\usepackage{booktabs}
\usepackage{hyperref}

\hypersetup{
  colorlinks=true,
  linkcolor=blue,
  citecolor=blue,
  urlcolor=blue,
  pdftitle={Spectral Stability, Fatou Basin Contraction, and Machine-Checked Multipliers in Holomorphic Extensions of Generalized Syracuse Dynamics},
  pdfauthor={Volkan Dagli, Zerrin Dagli, Daghan Dagli, Emrah Yurtseven, Murat Emre Ozkaya}
}

\newtheorem{theorem}{Theorem}[section]
\newtheorem{lemma}[theorem]{Lemma}

\theoremstyle{remark}

\title{\vspace{-1.0em}\textbf{\Large Spectral Stability, Fatou Basin Contraction, and Machine-Checked Multipliers in Holomorphic Extensions of Generalized Syracuse Dynamics}}

\author{
  \textbf{Volkan Da\u{g}l\i}$^{1,2,*}$,
  \textbf{Zerrin Da\u{g}l\i}$^{3}$,
  \textbf{Da\u{g}han Da\u{g}l\i}$^{4}$,
  \textbf{Emrah Yurtseven}$^{5}$,
  \textbf{Murat Emre \"{O}zkaya}$^{6}$\\[4pt]
  \small $^{1}$Anadolu University, Eski\c{s}ehir, T\"{u}rkiye \quad
  $^{2}$ITouch Systems, Adana, T\"{u}rkiye \quad
  $^{3}$Mersin University, Mersin, T\"{u}rkiye\\
  \small $^{4}$Toros Science High School, MEV (Toros University), Mersin, T\"{u}rkiye\\
  \small $^{5}$Near East University, Nicosia, Northern Cyprus \quad
  $^{6}$Marmara University, Istanbul, T\"{u}rkiye\\
  \small $^{*}$Corresponding author: \texttt{ask@answerr.me} \quad
  DOI: \href{https://doi.org/10.5281/zenodo.24060394}{10.5281/zenodo.24060394}
  (Concept DOI: \href{https://doi.org/10.5281/zenodo.24060393}{10.5281/zenodo.24060393})
}

\date{}

\begin{document}
\maketitle

\begin{abstract}
\noindent
We embed the Syracuse operator $T(n) = n/2$ (even $n$), $T(n) = (3n+1)/2$ (odd $n$) and the Collatz map $C(n)$ into $\mathbb{C}$ via entire functions of exponential type $\pi$: $f(z) = \frac{1}{4}(1 + 4z - (1 + 2z)\cos(\pi z))$ and $F(z) = \frac{1}{4}(2 + 7z - (2 + 5z)\cos(\pi z))$. We prove five results: (1) every integer periodic cycle $C \subset \mathbb{N}^+$ of $T$ is a strictly attracting Fatou cycle of $f$ with multiplier $\mu_f(C) = 3^{N_{\mathrm{odd}}}/2^N = \prod_{n \in C\text{ odd}} \frac{3n}{3n+1} \in (0,1)$, and $\mu_f(O_{1,2}) = \mu_F(O_{1,4,2}) = 3/4 < 1$; (2) on the Fatou basin $B(O_{1,2})$, $g = f^{\circ 2}$ conjugates to $\Phi_j(g(z)) = \frac{3}{4}\Phi_j(z)$ with Koenigs Taylor jets $c_2(1) = -\frac{13}{2}\pi^2$ and $d_2(2) = -\frac{9}{2}\pi^2$; (3) the Schwarzian derivative is strictly negative on $\mathbb{Z}$ ($S(f)(n) \le 3\pi^2 - \frac{3}{8}\pi^4 \approx -6.9196 < 0$), and a repelling Julia 2-cycle $O_{\mathrm{rep}} \subset J(f)$ ($\mu \approx 1.2148 > 1$) isolates $B(O_{1,2})$; (4) on $(\mathbb{Z}_2, \mu_{\mathrm{Haar}})$, the 2-adic Lyapunov exponent $\lambda_{\mathrm{Haar}} = \frac{1}{2}\ln(3/4) \approx -0.143841 < 0$ equals $\frac{1}{2}\ln|\mu_f(O_{1,2})|$; and (5) we certify 20 theorems in Lean 4 (\texttt{anc/CollatzAttractor.lean}, \texttt{0 sorry}). These results establish local Fatou invariants and 2-adic ergodic averages for $f(z)$, leaving open whether every $n_0 \in \mathbb{N}^+$ lies in $\bigcup_{k \ge 0} f^{-k}(B(O_{1,2}))$.
\end{abstract}

\section{Introduction}
Let $\mathbb{N}^+ = \{1, 2, 3, \dots\}$ denote the positive integers. The uncompressed Collatz operator $C : \mathbb{N}^+ \to \mathbb{N}^+$ and the compressed Syracuse operator $T : \mathbb{N}^+ \to \mathbb{N}^+$ are defined by
\begin{align}
  C(n) &= \begin{cases} n/2 & (n\text{ even}), \\ 3n+1 & (n\text{ odd}), \end{cases} \\
  T(n) &= \begin{cases} n/2 & (n\text{ even}), \\ (3n+1)/2 & (n\text{ odd}). \end{cases}
\end{align}
While computational verifications have confirmed convergence of $T^{\circ k}(n_0) \to 1$ up to $n_0 < 2^{71} \approx 2.36 \times 10^{21}$ \cite{Lagarias2010,Barina2021} and Tao \cite{Tao2022} established via logarithmic density that almost all orbits attain almost bounded values, the discrete parity indicator $(-1)^n$ prevents direct application of differential stability theory on $\mathbb{Z}$.

Following Chamberland \cite{Chamberland1996} and Letherman, Schleicher, and Wood \cite{LSW1999}, a natural analytical framework arises by lifting $T$ and $C$ to the complex plane $\mathbb{C}$ via entire functions of exponential type $\pi$. In this paper, we investigate the spectral geometry, higher-order Koenigs linearization jets, Schwarzian curvature, and 2-adic Haar ergodic resonance of these holomorphic extensions, and provide a complete machine-checked formalization of 20 foundational theorems in \textbf{Lean~4} (\texttt{0 sorry}).

\section{Canonical Entire Interpolations}
\begin{lemma}[Algebraic Equivalence and Integer Projection]\label{lem:interp}
Define the entire holomorphic maps $f, F : \mathbb{C} \to \mathbb{C}$ by
\begin{align}
  f(z) &= \frac{z}{2} + \frac{2z+1}{4}\bigl(1 - \cos(\pi z)\bigr) \nonumber\\
       &= \frac{1}{4}\bigl(1 + 4z - (1 + 2z)\cos(\pi z)\bigr), \label{eq:f}\\
  F(z) &= \frac{1}{4}\bigl(2 + 7z - (2 + 5z)\cos(\pi z)\bigr). \label{eq:F}
\end{align}
Then $f$ and $F$ are entire functions of order $\rho = 1$ and exponential type $\tau = \pi$ that restrict on $\mathbb{Z}$ to $f(n) = T(n)$ and $F(n) = C(n)$.
\end{lemma}
\begin{proof}
For $n = 2k \in 2\mathbb{Z}$, $\cos(2\pi k) = 1$, giving $f(2k) = k = T(2k)$ and $F(2k) = k = C(2k)$. For $n = 2k+1 \in 2\mathbb{Z}+1$, $\cos((2k+1)\pi) = -1$, giving $f(2k+1) = 3k+2 = (3n+1)/2 = T(n)$ and $F(2k+1) = 3(2k+1)+1 = C(n)$. These identities are machine-checked in \texttt{syracuse\_\allowbreak even\_\allowbreak projection}, \texttt{syracuse\_\allowbreak odd\_\allowbreak projection}, \texttt{chamberland\_\allowbreak even\_\allowbreak projection}, and \texttt{chamberland\_\allowbreak odd\_\allowbreak projection} in \texttt{CollatzAttractor.lean}.
\end{proof}

\begin{lemma}[Complex Derivative Jets on $\mathbb{Z}$]\label{lem:deriv}
The first three derivatives of $f(z)$ and the first derivative of $F(z)$ are
\begin{align}
  f'(z) &= 1 - \frac{1}{2}\cos(\pi z) + \frac{\pi}{4}(1+2z)\sin(\pi z),\\
  f''(z) &= \pi\sin(\pi z) + \frac{\pi^2}{4}(1+2z)\cos(\pi z),\\
  f'''(z) &= \frac{3\pi^2}{2}\cos(\pi z) - \frac{\pi^3}{4}(1+2z)\sin(\pi z),\\
  F'(z) &= \frac{7}{4} - \frac{5}{4}\cos(\pi z) + \frac{\pi}{4}(2+5z)\sin(\pi z).
\end{align}
At any integer $n \in \mathbb{Z}$, $\sin(\pi n) = 0$ and $\cos(\pi n) = (-1)^n$, so the transcendental sine terms vanish identically:
\begin{align}
  f'(2k) &= \frac{1}{2}, & f'(2k+1) &= \frac{3}{2},\\
  F'(2k) &= \frac{1}{2}, & F'(2k+1) &= 3.
\end{align}
\end{lemma}

\begin{table*}[t]
\centering
\caption{Complete Partition of Real Critical Points ($f'(c)=0$) and Periodic Orbits of $f(z)$ on $[0, 3]$.}\label{tab:partition}
\footnotesize
\begin{tabular}{@{}lllll@{}}
\toprule
\textbf{Dynamical Entity} & \textbf{Numerical Coordinates} & \textbf{Multiplier / Curvature} & \textbf{Classification} & \textbf{Captured Critical Point} \\
\midrule
Fixed Point $z_0 = 0$ & $0.0000000000$ & $\mu = f'(0) = 0.500000 < 1$ & Attracting Fatou fixed point & $c_0 \approx -0.1859909135$ \\
Repelling Fixed Point $x_1$ & $0.2773558495$ & $\mu = f'(x_1) \approx 1.495993 > 1$ & Julia basin boundary $\partial\mathcal{B}(0)$ & None ($\mathcal{J}(f)$ separator) \\
Integer 2-Cycle $\mathcal{O}_{1,2}$ & $\{1, 2\}$ & $\mu = g'(1) = 0.750000 < 1$ & Attracting integer Fatou cycle & $c_2 \approx 1.9583122157$ \\
Repelling 2-Cycle $\mathcal{O}_{\mathrm{rep}}$ & $\{1.02368612, 2.03342093\}$ & $\mu = g'(p_1) \approx 1.214780 > 1$ & Julia separator $\mathcal{J}(f)$ & None (separates $\mathcal{O}_{1,2}$ \& $\mathcal{O}^*$) \\
Irrational 2-Cycle $\mathcal{O}^*$ & $\{1.19253191, 2.13865634\}$ & $\mu = g'(q_1) \approx -0.712555$ & Secondary real Fatou 2-cycle & $c_1 \approx 1.2442606342$ \\
\bottomrule
\end{tabular}
\end{table*}

\section{Universal Cycle Attractiveness and Multiplier Cohomology}
\begin{theorem}[Universal Attractiveness of Integer Periodic Cycles]\label{thm:universal}
Let $\mathcal{C} = \{n_1, n_2, \dots, n_N\} \subset \mathbb{N}^+$ be any periodic cycle of the discrete Syracuse operator $T$, consisting of $N_{\mathrm{even}}$ even elements and $N_{\mathrm{odd}}$ odd elements ($N = N_{\mathrm{even}} + N_{\mathrm{odd}}$). Then $\mathcal{C}$ is unconditionally a strictly attracting periodic cycle of the entire map $f(z)$ in $\mathbb{C}$, and its holomorphic cycle multiplier satisfies the exact telescoping product identity:
\begin{equation}\label{eq:telescoping}
  \mu_f(\mathcal{C}) = \prod_{j=1}^N f'(n_j) = \frac{3^{N_{\mathrm{odd}}}}{2^N} = \prod_{\substack{n \in \mathcal{C}\\ n\text{ odd}}} \frac{3n}{3n+1} \in (0, 1).
\end{equation}
\end{theorem}
\begin{proof}
By Lemma~\ref{lem:deriv}, $f'(n_j) = 1/2$ if $n_j$ is even and $f'(n_j) = 3/2$ if $n_j$ is odd, so $\prod_{j=1}^N f'(n_j) = 3^{N_{\mathrm{odd}}}/2^N$. On the other hand, since $T(n_{j}) = n_{j+1}$ cyclically, multiplying the ratios $T(n)/n$ around the closed cycle telescopes to unity:
\begin{equation}
  1 = \prod_{j=1}^N \frac{T(n_j)}{n_j} = \left(\frac{1}{2}\right)^{N_{\mathrm{even}}} \prod_{\substack{n \in \mathcal{C}\\ n\text{ odd}}} \frac{3n+1}{2n} = \frac{1}{2^N} \prod_{\substack{n \in \mathcal{C}\\ n\text{ odd}}} \frac{3n+1}{n}.
\end{equation}
Multiplying both sides by $3^{N_{\mathrm{odd}}} \prod_{n\text{ odd}} \frac{n}{3n+1}$ yields \eqref{eq:telescoping}. Since $\frac{3n}{3n+1} \in [\frac{3}{4}, 1)$ for every $n \in \mathbb{N}^+$, $\mu_f(\mathcal{C}) \in (0,1)$, so $\mathcal{C}$ lies in the Fatou set $\mathcal{F}(f)$ as a strictly attracting basin. For the fundamental cycle $\mathcal{O}_{1,2} = \{1, 2\}$ ($N=2, N_{\mathrm{odd}}=1$), the single odd member $n=1$ gives $\mu_f(\mathcal{O}_{1,2}) = \frac{3}{4} = 0.75 < 1$. Furthermore, for the uncompressed cycle $\mathcal{O}_{1,4,2} = \{1, 4, 2\}$ of $F(z)$, Lemma~\ref{lem:deriv} gives $\mu_F(\mathcal{O}_{1,4,2}) = F'(1)F'(4)F'(2) = 3 \cdot \frac{1}{2} \cdot \frac{1}{2} = \frac{3}{4} = \mu_f(\mathcal{O}_{1,2})$ (machine-checked in \texttt{multiplier\_\allowbreak cohomology\_\allowbreak invariance}).
\end{proof}

\section{Closed-Form Koenigs Linearization Curvature Jets}
\begin{theorem}[Koenigs Linearization and Quadratic Curvature Jets]\label{thm:koenigs}
Let $g(z) = f(f(z))$ denote the second iterate of $f$, for which $z_1=1$ and $z_2=2$ are fixed points with multiplier $\mu = g'(1) = g'(2) = \frac{3}{4} \in (0,1)$. By the Koenigs Linearization Theorem \cite{Milnor2006,Koenigs1884}, there exist unique holomorphic coordinate charts $\Phi_j : U_j \to \mathbb{C}$ ($j \in \{1,2\}$) on neighborhoods $U_j \subset \mathcal{B}(\mathcal{O}_{1,2})$ of $z_j$ normalized by $\Phi_j(z_j)=0$ and $\Phi_j'(z_j)=1$ such that
\begin{equation}\label{eq:koenigs_eq}
  \Phi_j\bigl(g(z)\bigr) = \frac{3}{4}\Phi_j(z), \quad \forall z \in U_j.
\end{equation}
Expanding $\Phi_1(z) = (z-1) + c_2(z-1)^2 + O((z-1)^3)$ around $z_1=1$ and $\Phi_2(z) = (z-2) + d_2(z-2)^2 + O((z-2)^3)$ around $z_2=2$, the exact second derivatives of $g$ and quadratic Koenigs curvature coefficients are
\begin{align}
  g''(1) &= \frac{39}{16}\pi^2, & c_2 &= \frac{g''(1)}{2(\mu^2 - \mu)} = -\frac{13}{2}\pi^2 \approx -64.15, \\
  g''(2) &= \frac{27}{16}\pi^2, & d_2 &= \frac{g''(2)}{2(\mu^2 - \mu)} = -\frac{9}{2}\pi^2 \approx -44.41.
\end{align}
\end{theorem}
\begin{proof}
By the chain rule, $g''(z) = f''(f(z))(f'(z))^2 + f'(f(z))f''(z)$. Evaluating Lemma~\ref{lem:deriv} at $z=1$ and $z=2$ gives $f''(1) = -\frac{3}{4}\pi^2$ and $f''(2) = \frac{5}{4}\pi^2$. Thus
\begin{align*}
  g''(1) &= f''(2)\bigl(f'(1)\bigr)^2 + f'(2)f''(1) \\
         &= \frac{5\pi^2}{4}\cdot\frac{9}{4} - \frac{3\pi^2}{8} = \frac{39}{16}\pi^2,\\
  g''(2) &= f''(1)\bigl(f'(2)\bigr)^2 + f'(1)f''(2) \\
         &= -\frac{3\pi^2}{16} + \frac{15\pi^2}{8} = \frac{27}{16}\pi^2.
\end{align*}
Differentiating \eqref{eq:koenigs_eq} twice at $z_j$ yields $\Phi_j''(z_j)\mu^2 + g''(z_j) = \mu\Phi_j''(z_j)$, so $2c_2(\frac{9}{16}-\frac{3}{4}) + \frac{39}{16}\pi^2 = 0$, giving $c_2 = -\frac{13}{2}\pi^2$ and analogously $d_2 = -\frac{9}{2}\pi^2$. These identities are machine-checked in \texttt{second\_\allowbreak iterate\_\allowbreak curvature\_\allowbreak jets}, \texttt{koenigs\_\allowbreak quadratic\_\allowbreak coeff\_\allowbreak at\_1}, and \texttt{koenigs\_\allowbreak quadratic\_\allowbreak coeff\_\allowbreak at\_2}.
\end{proof}

\section{Universal Negative Schwarzian Derivative and Fatou--Julia Partition}
\begin{theorem}[Universal Negative Schwarzian Derivative on $\mathbb{Z}$]\label{thm:schwarzian}
The Schwarzian derivative $S(f)(z) = \frac{f'''(z)}{f'(z)} - \frac{3}{2}\bigl(\frac{f''(z)}{f'(z)}\bigr)^2$ is strictly negative at every integer $n \in \mathbb{Z}$:
\begin{align}
  S(f)(2k) &= 3\pi^2 - \frac{3}{8}(1+4k)^2\pi^4, \\
  S(f)(2k+1) &= -\pi^2 - \frac{1}{6}(3+4k)^2\pi^4.
\end{align}
In particular, $\max_{n \in \mathbb{Z}} S(f)(n) = S(f)(0) = 3\pi^2 - \frac{3}{8}\pi^4 \approx -6.9196 < 0$, and $S(f)(n) = -\mathcal{O}(n^2\pi^4) \to -\infty$ as $|n| \to \infty$.
\end{theorem}
\begin{proof}
At $n=2k$, Lemma~\ref{lem:deriv} gives $f'(2k)=1/2$, $f''(2k) = \frac{1+4k}{4}\pi^2$, and $f'''(2k) = \frac{3}{2}\pi^2$, whence $S(f)(2k) = 3\pi^2 - \frac{3}{8}(1+4k)^2\pi^4 \le 3\pi^2 - \frac{3}{8}\pi^4 < 0$. At $n=2k+1$, $f'(2k+1)=3/2$, $f''(2k+1) = -\frac{3+4k}{4}\pi^2$, and $f'''(2k+1) = -\frac{3}{2}\pi^2$, giving $S(f)(2k+1) = -\pi^2 - \frac{1}{6}(3+4k)^2\pi^4 < 0$.
\end{proof}

By Singer's Theorem \cite{Singer1978}, negative Schwarzian curvature prevents positive-Schwarzian inflection traps and ensures that every immediate attracting real basin contains at least one critical point $f'(c)=0$. Table~\ref{tab:partition} summarizes the real critical points and periodic orbits on $[0, 3]$, confirming that the immediate integer basin $\mathcal{B}(\mathcal{O}_{1,2})$ captures the critical point $c_2 \approx 1.9583122157$, while the repelling Julia 2-cycle $\mathcal{O}_{\mathrm{rep}} \approx \{1.02368612, 2.03342093\} \subset \mathcal{J}(f)$ ($\mu \approx 1.214780 > 1$) separates $\mathcal{B}(\mathcal{O}_{1,2})$ from Chamberland's irrational 2-cycle $\mathcal{O}^*$.

\section{The 2-Adic Haar--Fatou Resonance Identity}
\begin{theorem}[2-Adic Haar--Fatou Resonance Identity]\label{thm:haar}
The discrete Syracuse operator $T$ extends continuously to a measure-preserving ergodic shift on the compact ring of 2-adic integers $\mathbb{Z}_2$ equipped with its normalized Haar probability measure $\mu_{\mathrm{Haar}}$ ($\mu_{\mathrm{Haar}}(2\mathbb{Z}_2) = \mu_{\mathrm{Haar}}(2\mathbb{Z}_2+1) = 1/2$) \cite{BernsteinLagarias1996}. For $\mu_{\mathrm{Haar}}$-almost every 2-adic trajectory, the asymptotic holomorphic Lyapunov exponent equals one-half the logarithmic multiplier of the fundamental Fatou cycle $\mathcal{O}_{1,2}$:
\begin{align}
  \lambda_{\mathrm{Haar}}
  &= \int_{\mathbb{Z}_2} \ln\bigl|f'(\omega)\bigr|\,d\mu_{\mathrm{Haar}}(\omega)
   = \frac{1}{2}\ln\frac{1}{2} + \frac{1}{2}\ln\frac{3}{2} \nonumber\\
  &= \frac{1}{2}\ln\frac{3}{4}
   = \frac{1}{2}\ln\bigl|\mu_f(\mathcal{O}_{1,2})\bigr|
   \approx -0.143841 < 0.
\end{align}
Furthermore, a trajectory block of $N = N_{\mathrm{even}} + N_{\mathrm{odd}}$ steps contracts ($\lambda < 0$) if and only if $N_{\mathrm{even}}/N_{\mathrm{odd}} > \gamma_{\mathrm{crit}} = \ln(3/2)/\ln 2 \approx 0.5849625$.
\end{theorem}
\begin{proof}
By Lemma~\ref{lem:deriv}, $\ln|f'(\omega)| = -\ln 2$ on the clopen cylinder $2\mathbb{Z}_2$ and $\ln|f'(\omega)| = \ln(3/2)$ on $2\mathbb{Z}_2+1$. Since $T$ is conjugate to the 2-adic Bernoulli shift \cite{BernsteinLagarias1996}, Birkhoff's Ergodic Theorem yields
\begin{equation*}
  \lim_{N\to\infty}\frac{1}{N}\sum_{k=0}^{N-1}\ln\bigl|f'(T^{\circ k}(\omega))\bigr|
  = \frac{1}{2}\ln\frac{1}{2} + \frac{1}{2}\ln\frac{3}{2} = \frac{1}{2}\ln\frac{3}{4}
\end{equation*}
for $\mu_{\mathrm{Haar}}$-a.e.\ $\omega \in \mathbb{Z}_2$. The rational block contraction $2^{10}\cdot(1/2)^5(3/2)^5 = 243 < 1024$ is machine-checked in \texttt{haar\_\allowbreak ergodic\_\allowbreak block\_\allowbreak contraction}.
\end{proof}

Table~2 reports our high-throughput verification across $500,000$ consecutive integer trajectories ($7.46\,\mathrm{s}$ execution, $67,024$ orbits/s), demonstrating that as $N$ grows from $10^2$ to $5 \times 10^5$, the empirical parity ratio $N_{\mathrm{even}}/N_{\mathrm{odd}}$ approaches the 2-adic Haar ratio $1.0000$ from above ($1.0146 > \gamma_{\mathrm{crit}} \approx 0.5850$), while $\bar{\lambda}_N = -0.175737 < 0$ converges monotonically toward $\lambda_{\mathrm{Haar}} \approx -0.143841$.

\par\medskip\noindent
\begin{minipage}{\columnwidth}
\centering
\footnotesize
\textbf{Table 2:} Multi-Decade Convergence Toward the 2-Adic Haar Limit.\\[3pt]
\setlength{\tabcolsep}{3.8pt}
\begin{tabular}{@{}lrrrr@{}}
\toprule
\textbf{Range $[1, N]$} & \textbf{Mean Steps} & \textbf{Max Steps} & \textbf{$N_{\mathrm{even}}/N_{\mathrm{odd}}$} & \textbf{Mean $\bar{\lambda}_N$} \\
\midrule
$[1, 10^2]$ & $20.99$ & $73$ & $1.2398$ & $-0.223346$ \\
$[1, 10^3]$ & $39.25$ & $112$ & $1.0772$ & $-0.193207$ \\
$[1, 10^4]$ & $54.97$ & $164$ & $1.0457$ & $-0.184774$ \\
$[1, 10^5]$ & $69.34$ & $220$ & $1.0258$ & $-0.178962$ \\
$[1, 5\times 10^5]$ & $82.64$ & $281$ & $1.0146$ & $-0.175737$ \\
\midrule
\textbf{2-Adic Limit} & $\sim c\ln N$ & $\infty$ & \textbf{$1.0000$} & \textbf{$-0.143841$} \\
\bottomrule
\end{tabular}
\end{minipage}
\par\medskip

\section{Machine-Checked Formalization in Lean 4}
To eliminate symbolic and floating-point ambiguity, we formalized 20 algebraic, spectral, Koenigs curvature, and Diophantine theorems in \textbf{Lean~4} (\texttt{anc/CollatzAttractor.lean}) with \textbf{zero unproven assumptions (\texttt{0 sorry})}, summarized in Table~3.

\par\medskip\noindent
\begin{minipage}{\columnwidth}
\centering
\footnotesize
\textbf{Table 3:} Summary of the 20 Machine-Checked Theorems in \texttt{CollatzAttractor.lean} (\texttt{0 sorry}).\\[3pt]
\setlength{\tabcolsep}{3.5pt}
\begin{tabular}{@{}lll@{}}
\toprule
\textbf{Sec.} & \textbf{Verified Lean 4 Theorem Modules} & \textbf{Status} \\
\midrule
\S 1 & \texttt{syracuse\_even/odd}, \texttt{chamberland\_even/odd} & \texttt{0 sorry} \\
\S 2 & \texttt{df\_syracuse}, \texttt{multiplier\_cohomology} & \texttt{0 sorry} \\
\S 3 & \texttt{second\_iterate\_jets}, \texttt{koenigs\_coeff\_1/2} & \texttt{0 sorry} \\
\S 4 & \texttt{haar\_ergodic\_block}, \texttt{ten\_cycle\_damping} & \texttt{0 sorry} \\
\bottomrule
\end{tabular}
\end{minipage}
\par\medskip

\section*{Acknowledgments}
The authors thank Bekirhan Da\u{g}l\i{} and ITouch Systems Research \& Development (Cukurova University Science Park, Adana, T\"{u}rkiye) for dedicated computing infrastructure. Companion replication archive available at \href{https://zenodo.org/records/24060394}{doi:10.5281/zenodo.24060394} (Concept DOI: \href{https://doi.org/10.5281/zenodo.24060393}{10.5281/zenodo.24060393}).

\end{document}